\documentclass[journal]{IEEEtran}

\usepackage{url}
\usepackage{hyperref}
\usepackage[hyphenbreaks]{breakurl}
\usepackage[pdftex]{graphicx}
\usepackage{xcolor}
   \definecolor{mygreen}{RGB}{51, 153, 51}
   \definecolor{blue_gray}{RGB}{102, 153, 204}
\usepackage{multirow}
\usepackage{amsmath}
\usepackage{mathrsfs}
\usepackage{amsfonts,amssymb}
\usepackage{booktabs}

\begin{document}
\title{COVID-19 CT Image Synthesis with a Conditional Generative Adversarial Network}

\author{Yifan Jiang,~\IEEEmembership{Member,~IEEE,}
        Han Chen, Murray Loew,~\IEEEmembership{Fellow,~IEEE}
        and Hanseok Ko,~\IEEEmembership{Senior Member,~IEEE}% <-this % stops a space
\thanks{This research work is supported by a National Research Foundation (NRF) grant funded by the MSIP of Korea (number 2019R1A2C2009480). (Corresponding author: Hanseok Ko.) Y. Jiang, H. Chen, and H. Ko are with the School of Electrical Engineering, Korea University, Seoul 02841, South Korea (e-mail:  yfjiang@ispl.korea.ac.kr;hanchen@ispl.korea.ac.kr; hsko@korea.ac.kr). M. Loew is with School of Engineering \& Applied Science, George Washington University, Washington, DC 20052, United States (e-mail:  loew@gwu.edu).}}

\markboth{JIANG \MakeLowercase{\textit{et al.}}: COVID-19 CT IMAGE SYNTHESIS WITH A CONDITIONAL GENERATIVE ADVERSARIAL NETWORK}%
{JIANG \MakeLowercase{\textit{et al.}}: COVID-19 CT IMAGE SYNTHESIS WITH A CONDITIONAL GENERATIVE ADVERSARIAL NETWORK}

\maketitle

\begin{abstract}
Coronavirus disease 2019 (COVID-19) is an ongoing global pandemic that has spread rapidly since December 2019. Real-time reverse transcription polymerase chain reaction (rRT-PCR) and chest computed tomography (CT) imaging both play an important role in COVID-19 diagnosis. Chest CT imaging offers the benefits of quick reporting, a low cost, and high sensitivity for the detection of pulmonary infection. Recently, deep-learning-based computer vision methods have demonstrated great promise for use in medical imaging applications, including X-rays, magnetic resonance imaging, and CT imaging. However, training a deep-learning model requires large volumes of data, and medical staff faces a high risk when collecting COVID-19 CT data due to the high infectivity of the disease. Another issue is the lack of experts available for data labeling. In order to meet the data requirements for COVID-19 CT imaging, we propose a CT image synthesis approach based on a conditional generative adversarial network that can effectively generate high-quality and realistic COVID-19 CT images for use in deep-learning-based medical imaging tasks. Experimental results show that the proposed method outperforms other state-of-the-art image synthesis methods with the generated COVID-19 CT images and indicates promising for various machine learning applications including semantic segmentation and classification.
\end{abstract}


\IEEEpeerreviewmaketitle

\section{Introduction}

\IEEEPARstart{C}{oronavirus} disease 2019 (COVID-19) \cite{covid19}, which was first identified in Wuhan, China, in December 2019, was declared a pandemic in March 2020 by the World Health Organization (WHO). As of 21 July, there had been more than 14 million confirmed cases and 609,198 deaths across 188 countries and territories \cite{covid19_num}. COVID-19 is the result of severe acute respiratory syndrome coronavirus 2 (SARS-CoV-2), and its most common symptoms include fever, dry cough, a loss of appetite, and fatigue, with common complications including pneumonia, liver injury, and septic shock \cite{guan2020clinical, covid19_test}.

There are two main diagnostic approaches for COVID-19: rRT-PCR and chest computed tomography (CT) imaging \cite{covid19_test}. In rRT-PCR, an RNA template is first converted by reverse transcriptase into complementary DNA (cDNA), which is then used as a template for exponential amplification using polymerase chain reaction (PCR). However, the sensitivity of rRT-PCR is relative low for COVID-19 testing \cite{ai2020correlation, fang2020sensitivity}. As an alternative, chest CT scans can be used to take tomographic images from the chest area at different angles with post-computed X-ray measurements. This approach has a higher sensitivity to COVID-19 and is less resource-intensive than traditional rRT-PCR \cite{ai2020correlation, fang2020sensitivity}.

Over time, artificial intelligence (AI) has come to play an important role in medical imaging tasks, including CT imaging \cite{yan2019holistic, cui2019toothnet}, magnetic resonance imaging (MRI) \cite{zhang2019reducing} and X-ray imaging \cite{ying2019x2ct}. Deep learning is a particularly powerful AI approach that has been successfully employed in a wide range of medical imaging tasks due to the massive volumes of data that are now available. These large datasets allow deep-learning networks to be well-trained, extending their generalizability for use in various applications. However, the collection of COVID-19 data for use in deep-learning models is far more difficult than normal data collection. Because COVID-19 is highly contagious \cite{covid19_test}, medical staff require full-length protection for CT scans, and the CT scanner and other equipment need to be carefully disinfected after an operation. In addition, certain tasks, such as CT image segmentation, require well-labeled data, which is labor-intensive. These problems mean that the COVID-19 CT data collection process can be difficult and time-consuming.

\begin{figure}[!ht]
\centering
\includegraphics[width=8cm]{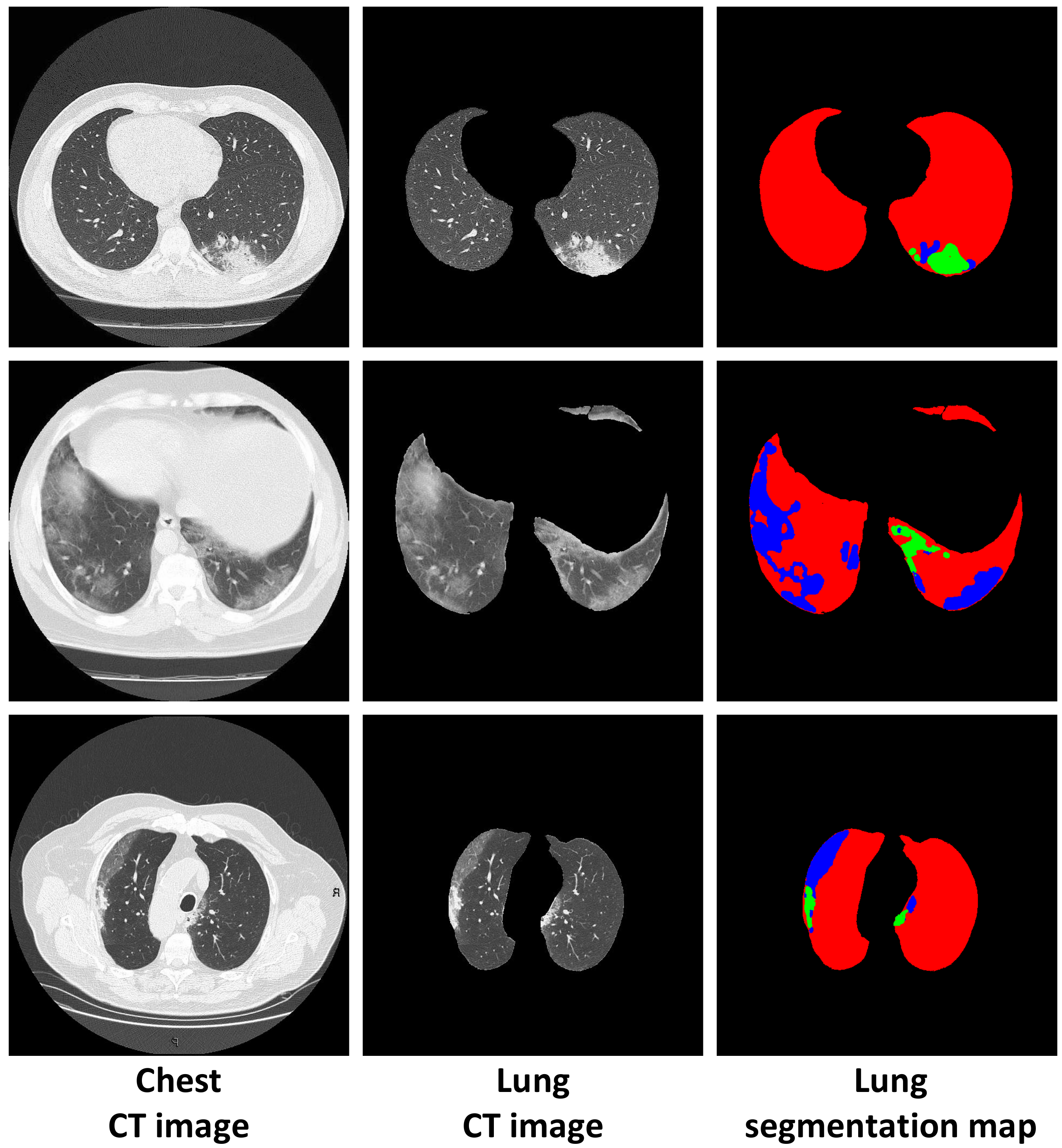}
\caption{Example CT images from three COVID-19 patients. The first column shows CT images of the entire chest, the second column contains CT images of the lungs only, and the third column shows the corresponding segmentation map, with the lung region colored red, ground-glass opacity colored blue, and areas of consolidation colored green.}
\label{fig1} % this gives the figure a unique name that you can refer to in the main text \ref{fig:...}
\end{figure}

In order to speed up the COVID-19 CT data collection process for deep-learning-based CT imaging and to protect medical personnel from possible infection when coming into contact with COVID-19 patients, we propose a novel cGAN structure which contains a global-local generator and a multi-resolution discriminator. Both the above generator and discriminator are dual network design so that they can learn global and local information of CT images individually. Also, the dual structure has a communication mechanism for information exchange so that it helps to generate a realistic CT image with both stable global structure and diverse local details. The main contributions of the proposed method are as follows:

\begin{itemize}

\item [(1)] We presented a dual generator structure (global-local generator). This dual global-local generator contains two individual generators that address and reflect different-level of information from CT data.
\item [(2)]	We proposed a dual discriminator (multi-resolution discriminator) that contains two sub-discriminators. These two discriminators learn to distinguish input from real or fake. They are specially designed for learning from full-resolution CT data and half-resolution ones, respectively.
\item [(3)] A dynamic communication mechanism is proposed for both generator and discriminator. In the case of the generator, a dynamic element-wise sum process (DESUM) helps generators balance the information of the lung area and small lesion area by dynamically weighting two terms during the element-wise sum process.  It also prevents the generator from overweighting details like a traditional cGAN model does for wild scene dataset. For the discriminator, a dynamic feature matching process (DFM) is proposed for dynamically weighting the loss terms from two inputs with different resolutions. In particular, it allows the half-resolution discriminator to receive more information with lung structure or large lesion area. It also offers more features of small lesion area to full-resolution discriminator. This dual multi-resolution discriminator helps to stabilize the training process and improves the image quality of the synthetic data.
\item [(4)] The proposed method outperforms other state-of-the-art image synthesizers in several image-quality metrics and demonstrates its potential for use in image synthesis for computer vision tasks such as semantic segmentation for COVID-19 chest CT imaging.    
\item [(5)] A safe COVID-19 chest CT data collection method based on image synthesis is presented. The potential applications of proposed method are summarized as follows: (a) COVID-19 CT synthesis method can be applied to the data augmentation task for the deep learning based COVID-19 diagnosis approaches; (b) COVID-19 CT synthesis method can also be utilized to train the intern radiologists who may need abundant snapshots of COVID-19 CT scans for training purposes; (c) The proposed COVID-19 CT synthesis method can be easily transferred from CT imaging domain to another medical imaging area (e.g. X-ray, MRI).

\end{itemize}

\section{Related works}

\noindent \textbf{Generative adversarial networks.} Generative adversarial networks (GANs) were first reported in 2014 \cite{goodfellow2014generative}, and they have since been widely applied to many practical applications, including image synthesis \cite{isola2017image,wang2018high,park2019semantic,zhu2019sean},  image enhancement \cite{chen2018deep,wan2020bringing}, human pose estimation \cite{yang20183d,ma2017pose}, and video generation \cite{wang2018video,wang2019few}. A GAN structure generally consists of a generator and a discriminator, where the goal of the generator is to fool the discriminator by generating a synthetic sample that cannot be distinguished from real samples. A common GAN extension is the conditional generative adversarial network (cGAN) \cite{mirza2014conditional}, which generates images that are conditional on class labels. cGAN always produces more realistic results than traditional GANs due to the extra information from these conditional labels.

\noindent \textbf{Conditional image-to-image translation.}  Conditional image- to-image translation methods can be divided into three categories based on the input conditions. Class-conditional methods take class-wise labels as input to synthesize images \cite{mirza2014conditional,odena2017conditional,caesar2018coco,mescheder2018training} while, more recently, text-conditional methods have been introduced \cite{xu2018attngan,hong2018inferring}. cGAN-based methods \cite{isola2017image,wang2018high,park2019semantic,zhu2019sean,zhu2017unpaired,zhu2017toward,liu2017unsupervised,huang2018multimodal,hong2018inferring,xu2018attngan,zhao2019image} have been widely used for various image-to-image translation methods, including unsupervised \cite{liu2017unsupervised}, high-quality \cite{wang2018high}, multi-modal \cite{zhu2017unpaired, park2019semantic, zhu2019sean}, and semantic layout conditional image-to-image translation \cite{isola2017image,wang2018high,park2019semantic,zhu2019sean}. In semantic layout conditional methods, realistic images are synthesized under the navigation of the semantic layout, meaning that it is easier to control a particular region of the image.

\begin{figure*}[!ht]
\centering
\includegraphics[width=18cm]{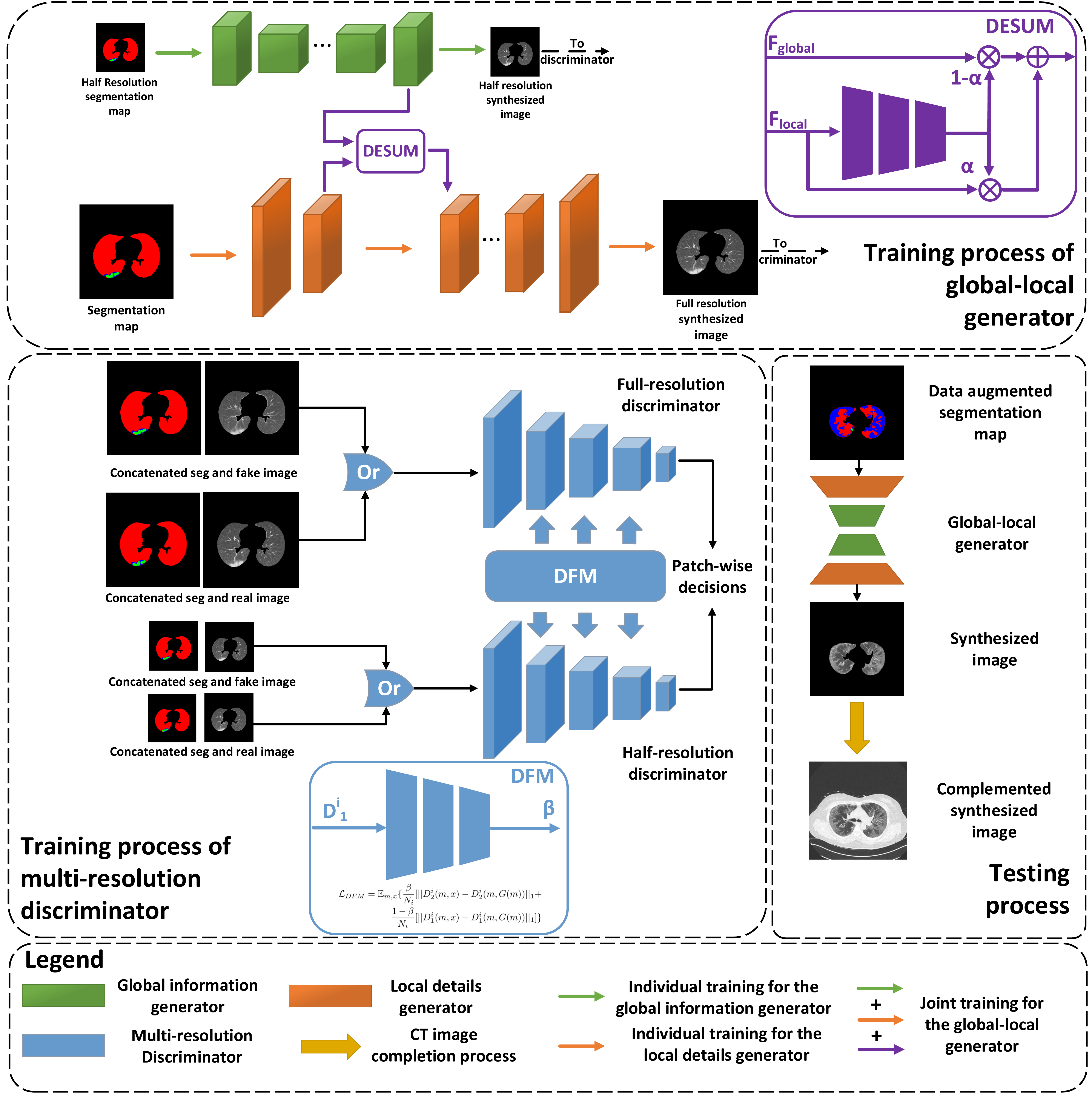}
\caption{Overview of the proposed method. The upper section containing the training process of the global-local generator and multi-resolution discriminator, while the lower right section shows the testing process. Within the global-local generator blocks, two types of generator are present: a \textcolor{orange}{global information generator} and \textcolor{mygreen}{local detail generator}. Two individual training processes and single-joint training process are depicted in three different colorized arrows. DESUM block represents the dynamic element-wise sum process which is shown in purple. The \textcolor{blue_gray}{multi-resolution discriminator} is depicted in blue. And the dynamic feature matching process (DFM) is also shown as a blue block. The synthesized images are transferred from the generator to the discriminator, and this process is shown as the dashed arrow. The yellow arrow shows the completion step for the process in which the non-lung region for the synthesized lung image is added.}
\label{fig2} % this gives the figure a unique name that you can refer to in the main text \ref{fig:...}
\end{figure*}

\noindent \textbf{AI-based diagnosis using COVID-19 CT imaging.} Since the outbreak of COVID-19, many researchers have turned to CT imaging technology in order to diagnose and investigate this disease. COVID-19 diagnosis methods based on chest CT imaging have been introduced in order to improve test efficiency \cite{kang2020diagnosis,zhang2020clinically,ardakani2020application,li2020artificial}. Rather than using CT imaging for rapid COVID-19 diagnosis, semantic segmentation approaches have been employed to clearly label the focus position in order to make it easier for medical personnel to identify infected regions in a CT image \cite{zhou2020automatic,xie2020relational,voulodimos2020deep,chen2020residual,fan2020inf}. As an alternative to working at the pixel-level, high-level classification or detection approaches have been proposed \cite{zheng2020deep,gozes2020coronavirus,hu2020weakly}, which can allow medical imaging experts to rapidly locate areas of infection, thus speeding up the diagnosis process. Though two CT image synthesis methods have been previously reported \cite{lauritzen2019evaluation,chen2020novel}, they did not focus on COVID-19 or lung CT imaging. cGAN was introduced to COVID-19 CT image synthesis task firstly by \cite{liu20203d}, which transforms a normal 3D CT slice to an abnormally synthetic slice under the condition of 3D noise. 

\section{COVID-19 CT Image Synthesis with a Conditional Generative Adversarial Network}

In this paper, we propose a cGAN-based COVID-19 CT image synthesis method. Here, COVID-19 CT image synthesis is formulated as a semantic-layout-conditional image-to-image translation task. The structure consisting of two main components: a global-local generator and a multi-resolution discriminator. During the training stage, the semantic segmentation map of a corresponding CT image is passed to the global-local generator, where the label information from the segmentation map is extracted via down-sampling and re-rendered to generate a synthesized image via up-sampling. The segmentation map is then concatenated with the corresponding CT image or synthesized CT image to form the input for the multi-resolution discriminator, which is used to distinguish the input as either real or synthesized. The decisions from the discriminator are used to calculate the loss and update the parameters for both the generator and discriminator. During the testing stage, only the generator is involved. A data augmented segmentation map is used as input for the generator, from which a realistic synthesized image can be obtained after extraction and re-rendering. This synthesized lung CT image is then combined with the non-lung area to form a completely synthesized CT image as the final result. Figure \ref{fig2} presents an overview of the proposed method.

\subsection{Global-local generator}

The global-local generator $G$ is a dual network which has two sub-components: global-information generator $G_1$ and local-detail generator $G_2$. These generators work together by moving in a coarse-to-fine direction. $G_1$ takes charge of learning and re-rendering global information, which always contains high-level knowledge (e.g., semantic segmentation labels and image structure information). $G_2$ is then used for detail enhancement (e.g., image texture and fine structures).

We train the global-local generator using a three-step process:

\begin{figure}[!ht]
\centering
\includegraphics[width=8cm]{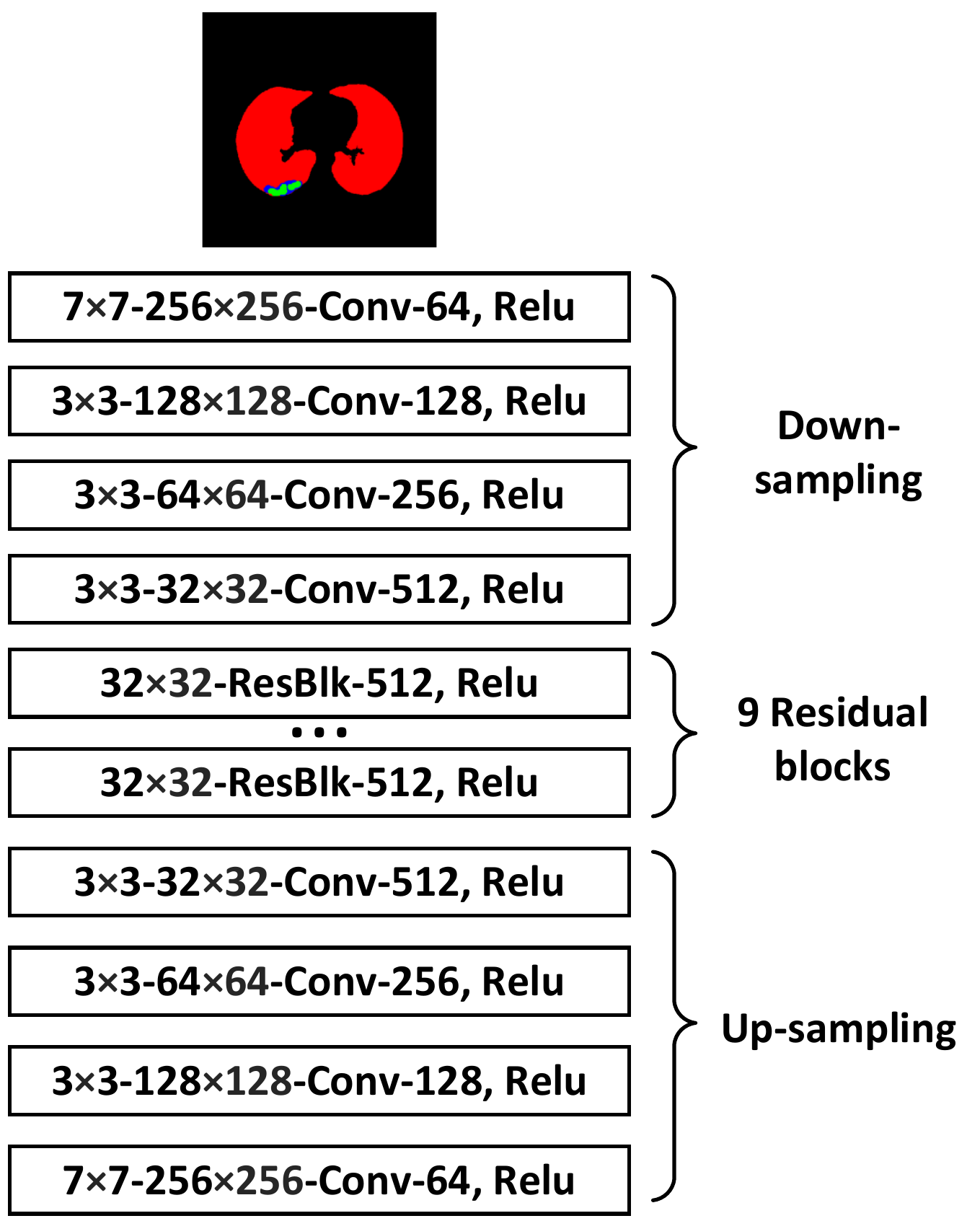}
\caption{The network structure of the global information generator $G_1$. The parameters of each layer are separated by the notation '-', e.g. for the first layer, $7\times7$ means the kernel size is 7, $256\times256$ is the size of the feature, Conv denotes the category of the layer, 64 is the channel number, and Relu is the activation function.}
\label{fig3} % this gives the figure a unique name that you can refer to in the main text \ref{fig:...}
\end{figure}

\subsubsection{Individual training for the global information generator}
The training process for $G$ starts with the training of the global information generator $G_1$. As shown in Figure \ref{fig3}, $G_1$ takes a half-resolution ($256\times 256$) segmentation map as input, which is then sent for down-sampling to reduce the feature dimensions to $32\times 32$. Nine residual blocks that maintain the dimensions at $32\times 32$ are used to reduce the computational complexity and generate a large reception field. Finally,  the features are up-sampled and reconstructed back into a half-resolution ($256\times 256$) synthesized image.

\begin{figure}[!ht]
\centering
\includegraphics[width=8cm]{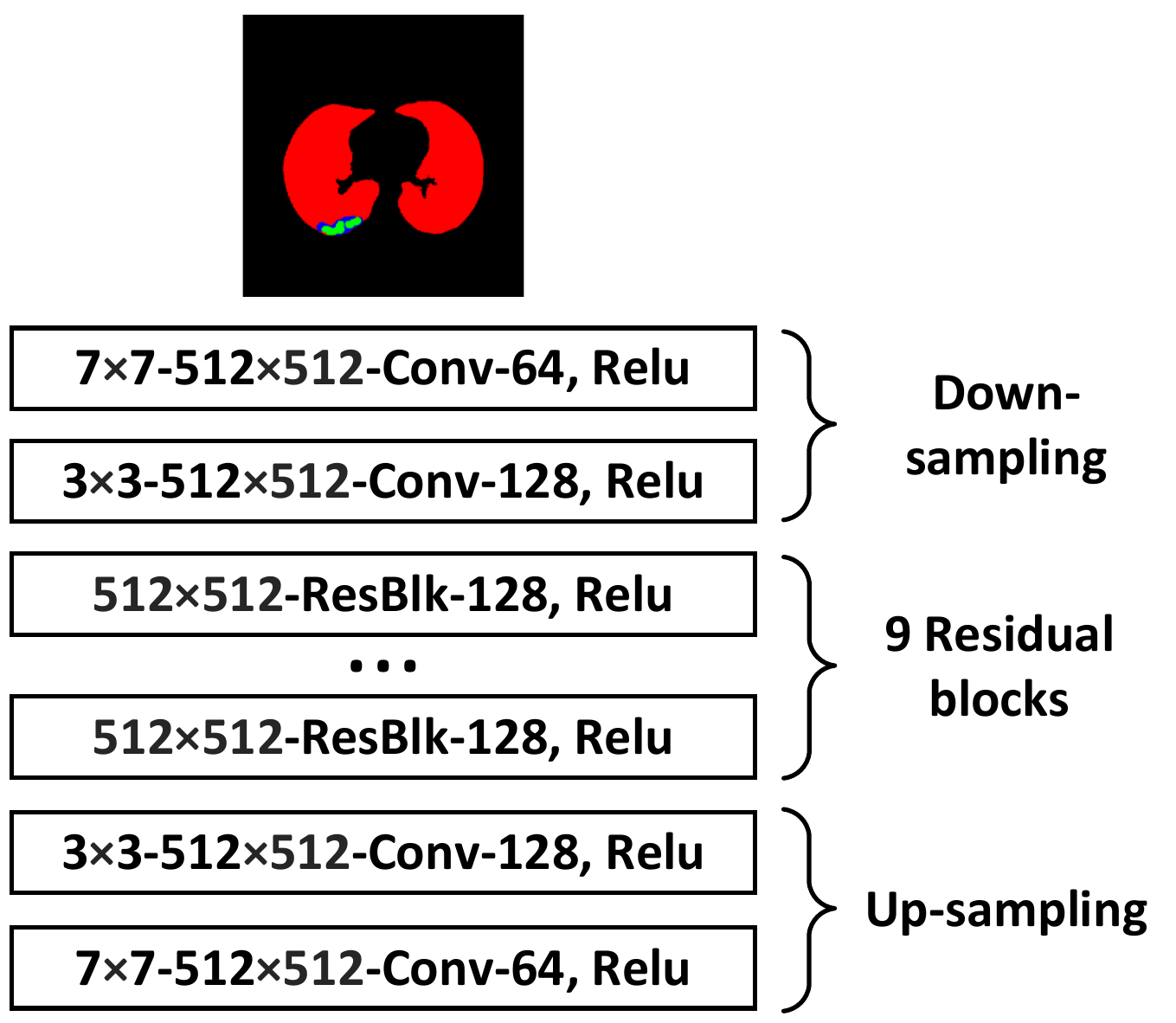}
\caption{Network structure of local details generator $G_2$.}
\label{fig4} % this gives the figure a unique name that you can refer to in the main text \ref{fig:...}
\end{figure}

\subsubsection{Individual training for the local details generator}
The structure of the local detail generator $G_2$,  which is similar to the structure of $G_1$, is shown in Figure \ref{fig4}. Rather than taking a low-resolution segmentation map as input, the local detail generator begins the synthesis process with a full-resolution segmentation map ($512\times 512$) and maintains this size throughout. That allows the local detail generator to fully learn the fine texture and structure and focus on low-level information within the input image. $G_2$ has a similar encoding-decoding training procedure as G1, though the output synthesized image is $512\times 512$.

\subsubsection{Joint training for the global-local generator}
After training $G_1$ and $G_2$ separately, a joint training process is conducted. This is shown in the global-local generator region of Figure \ref{fig2}. In the joint training stage, both $G_1$ and $G_2$ take the same input but with different resolutions (half- and full-resolution, respectively). The two networks run a forward process that differs from the individual training stage in which the residual blocks in $G_2$ takes the dynamic element-wise sum from the output feature maps from the up-sampling process in $G_1$ and the output feature maps from down-sampling in $G_2$, meaning that $G_2$ receives both global and local information to reconstruct the output.

This training strategy enables the global-local generator $G$ to effectively learn both global information and local details while also stabilizing the training process by simplifying it into three relatively simple procedures.

\begin{figure}[!ht]
\centering
\includegraphics[width=6cm]{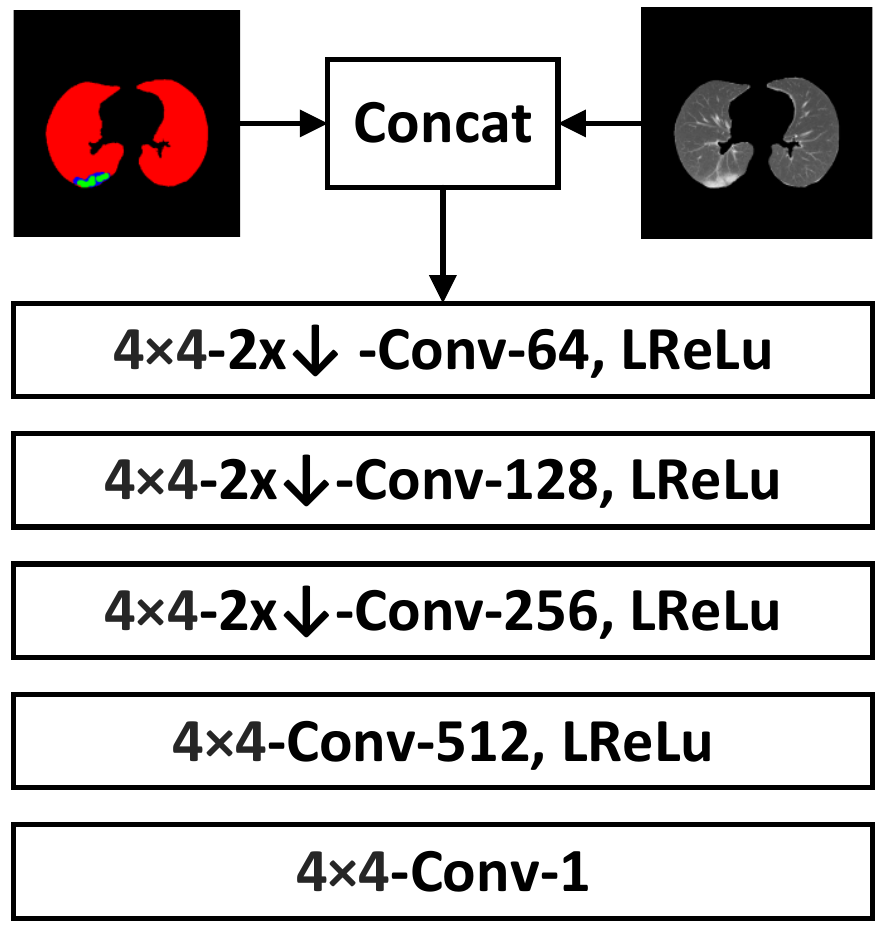}
\caption{Network structure of the multi-resolution discriminator $D$. $2\times \downarrow$ denotes down-sampling with a factor of 2.}
\label{fig5} % this gives the figure a unique name that you can refer to in the main text \ref{fig:...}
\end{figure}

\subsection{Multi-resolution discriminator}
A multi-resolution discriminator $D$ is proposed in this paper. This dual network structure consists of two sub-components: the full-resolution discriminator $D_1$ and the half-resolution discriminator $D_2$. We design above two discriminators by following the PatchGAN discriminator \cite{isola2017image}, therefore, proposed discriminator contains two PatchGAN discriminators. Two discriminators make patch-wise decisions rather than making a decision for the whole image. $D_1$ takes the full-resolution input and learns the local information from CT image, on the other hand, $D_2$ takes half-resolution image as input and focuses on the global information from CT image. In addition, we proposed a dynamic feature matching process (DFM) for improving the communication quality between $D_1$ and $D_2$ during the training process. As shown in Figure \ref{fig5}, we first down-sample the segmentation map and real image into half-resolution form, then the synthesized and real images are randomly chosen to be concatenated with the segmentation map to form two inputs (full- and half-resolution) for $D$. The target discriminator takes corresponding input and makes decision through a $70\times70$ receptive field, and the decision is represented as a decision matrix, which represents the patch-wise decisions for corresponding inputs. Then, patch-wise decisions are used to update $D$ with the DFM in order to adaptively share multiple resolution intermediate features between two discriminators.

The dual-discriminator design and dynamic feature matching process enable the multi-resolution discriminator $D$ to effectively learn local details, which can significantly improve the quality of the synthesized image. By assigning local and global discrimination to individual discriminators $D_1$ and $D_2$, and dynamically weighting the feature matching losses from $D_1$ and $D_2$, the global structure can be maintained while also enhancing the details of the synthesized images.

\subsection{Dynamic communication mechanism}
\subsubsection{Dynamic element-wise sum process}

The dynamic element-wise sum (DESUM) process is utilized in the joint training step of global-local generator. As shown in Figure 2, DESUM process takes two feature maps ($F_{global}$, $F_{local}$) from $G_1$ and $G_2$, respectively. We trained a weighting network that contains three convolutional layers and 2 fully-connected layers to dynamically compute the weight of two input terms from $G_2$. The DESUM process can be formulated as follow:

\begin{equation}
\begin{aligned}
F_{out} = \alpha F_{local} \oplus (1-\alpha) F_{global}
\end{aligned}
\label{eq0}
\end{equation}
where $\alpha$ is learned by the weighting network. This weighting network is updated during the joint training step. The DESUM process effectively helps to dynamically adjust the generator to specific input and balance the attention from global information to local details. To be specific, the DESUM process can dynamically weight more to $F_{local}$ when receiving an input that contains complex lesion area. On the other hand, the DESUM process is able to avoid the generator from overweight on tiny lesions but ignore global lung structure.

\subsubsection{Dynamic feature matching process}
As shown in Figure \ref{fig2}, the dynamic feature matching process (DFM) computes weight parameter $\beta$ for the dynamic feature matching loss ($\mathcal{L}_{DFM}$) which will be discussed in sub-section D. Similar to DESUM, DFM uses a CNN structure to calculate the weight parameter by observing an intermediate feature from $D_1$. However, DFM works on loss level rather than the feature level. By applying the DFM process to DFM loss, multi-resolution discriminator $D$ is able to balance between two resolution inputs and communicate with each other. Since the weight parameter $\beta$ is decided by intermediate feature $D_{1}^{i}$, which is the intermediate feature from $i^{th}$ layer of full-resolution discriminator $D_1$, the DFM network can obtain enough information of full-resolution input and weight $\mathcal{L}_{DFM}$ correctly.

\subsection{Learning objective}

The overall learning objective of proposed approach can be represented by equation \eqref{eq1}:

\begin{equation}
\begin{aligned}
\min_{G}\left[\max_{D_{1},D_{2}}\sum_{i=1}^{2}\mathcal{L}_{cGAN}\{G(m),D_i(m,x),D_i(m,G(m))\}+ \right.
\\
\phantom{=\;\;}
\left.\lambda\sum_{i=1}^{2}\mathcal{L}_{DFM}\{G(m),D_i(m,x),D_i(m,G(m))\}\right]
\end{aligned}
\label{eq1}
\end{equation}
There are two main loss terms in the overall learning objective function \eqref{eq1}: the loss for the cGAN $\mathcal{L}_{cGAN}$ and the loss for dynamic feature matching $\mathcal{L}_{DFM}$. The variable $x$ is the real input image and $m$ is the corresponding segmentation map. $G$ represents global-local generator while $D_i$ represents the full-resolution discriminator $D_1$ or half-resolution discriminator $D_2$. $G(m)$ denotes the synthesized image produced by generator $G$ with input segmentation map $s$, $D_i(m,x)$ and $D_i(m,G(m))$ are the patch-wise decisions made by multi-resolution discriminator $D$ with the real image or synthesized image as input, respectively. $\lambda$ is the weight factor of feature matching loss term.

We designed the cGAN loss function based on pix2pix \cite{isola2017image}, as shown in \eqref{eq2}

\begin{equation}
\begin{aligned}
\mathcal{L}_{cGAN}=\mathbb{E}_{m,x}[logD(m,x)]+\mathbb{E}_{m,x}[log(1-D(m,G(m)))]
\end{aligned}
\label{eq2}
\end{equation}
This loss term allows cGAN to generate a realistic synthesized image that can fool discriminator under the condition of the input segmentation map.

In order to help to improve the communication efficiency between multi-resolution discriminator $D_1$ and $D_2$, we proposed a dynamic feature matching loss (Eq. \eqref{eq3}) which is inspired by the feature matching loss from ref \cite{johnson2016perceptual}:
\begin{equation}
\begin{aligned}
\mathcal{L}_{DFM}=\mathbb{E}_{m,x}\sum_{i=1}^{3}\{\frac{\beta}{N_{i}}[||D_{2}^{i}(m,x)-D_{2}^{i}(m,G(m))||_1 + \\ \frac{1-\beta}{N_{i}}[||D_{1}^{i}(m,x)-D_{1}^{i}(m,G(m))||_1]\}
\end{aligned}
\label{eq3}
\end{equation}
where $i$ represents the $i^{th}$ layer of $D$ and $N_i$ is the total number of elements in the $i^{th}$ layer. $\beta$ is a weight parameter which is computed by dynamic feature matching process (described in sub-section C). Original feature matching loss only considers to manage the feature map difference between different layers within a single discriminator. In order to overcome the communication problem between two discriminators, the dynamic feature matching loss (DFM loss) dynamically weights the feature matching losses from the full- and half-resolution discriminators through observing an intermediate feature $D_{1}^{i}$. By applying DFM loss, it allows us to train $D_1$ and $D_2$ synchronously, and to learn the details from the inputs with different resolutions effectively.

\subsection{Testing process}

Rather than using both global-local generator $G$ and multi-resolution discriminator $D$ as in the training stage, we only utilize the pre-trained $G$ in the testing process. The input for $G$ in this stage is a data augmented segmentation map from the real data. During the practical deployment, the segmentation maps can be obtained by augmenting the segmentation maps which are made by experienced radiologists using standard image editing software. After passing it through $G$, a synthesized CT image of the lung area is generated. The final step in the process combines the synthesized lung image with the corresponding non-lung area from the real image to produce a complete synthesized image.

\section{Experiments}

\subsection{Experimental settings}

\noindent \textbf{Dataset.} In order to evaluate the proposed method and compare its performance to other state-of-the-art methods, we use 829 lung CT slices from nine COVID-19 patients, which were made public on 13 April 2020, by Radiopaedia \cite{covid19_data}. This dataset includes the original CT images, lung masks, and COVID-19 infection masks. The infection masks contain ground-glass opacity and consolidation labels, which are the two most common characteristics used for COVID-19 diagnosis in lung CT imaging \cite{chung2020ct}. In this experiment, we select 446 slices that contained the areas of infection. We divide the selected dataset into three parts: a training set for image synthesis task (300), a test set for image synthesis task (73), a test set for semantic segmentation task (73). To fully train the deep-learning-based model, data augmentation pre-processing is applied (Table \ref{table1}). The training set for the semantic segmentation tasks consists of real data and synthetic data: the real data comes from the test set of image synthesis task and the synthetic data is generated from the segmentation maps from the test set of image synthesis task.

\begin{table}[htbp] 
 \centering
 \caption{ORGANIZATION OF THE COVID-19 CT IMAGE DATASET} 
 \begin{tabular}{c  c  c } 
  \toprule 
  Dataset & Original count & After data augmentation \\ 
  \midrule 
 Training set \\ (image synthesis) & 300 & 12,000 \\ 
 Test set \\ (image synthesis) & 73 & 10,220 \\ 
 Training set \\ (semantic segmentation) & - & - \\
 Test set \\ (semantic segmentation) & 73 & 10,220 \\ 
  \bottomrule 
 \end{tabular} 
\label{table1}
\end{table}
The data augmentation methods include random resizing and cropping, random rotation, Gaussian noise, and elastic transform.

\begin{table*}[]
	\centering
	\caption{IMAGE QUALITY EVALUATION RESULTS OF SYNTHETIC CT IMAGES \hspace{\textwidth} (The best evaluation score is marked in bold. $\uparrow$ means higher number is better, and $\downarrow$ indicates lower number is better.)}
	\begin{tabular}{|c|c|c|c|c|c|c|c|c|}
		\hline
            Categories & \multicolumn{4}{|c|}{Complemented images} & \multicolumn{4}{|c|}{Lung only images}  \\ 
		\hline
		    Metrics & FID ($\downarrow$) & PSNR ($\uparrow$) & SSIM ($\uparrow$) & RMSE ($\downarrow$) & FID ($\downarrow$) & PSNR ($\uparrow$) & SSIM ($\uparrow$) & RMSE ($\downarrow$) \\
		\hline
		    OURS & \textbf{0.0327} & \textbf{26.89} & \textbf{0.8936} & \textbf{0.0813} & 0.3641 & \textbf{28.17} & \textbf{0.8959} & \textbf{0.2747} \\
		\hline
		    SEAN \cite{zhu2019sean} & 0.0341 & 26.69 & 0.8922 & 0.0837 & \textbf{0.3575} & 28.02 & 0.8952 & 0.2795 \\
		\hline
    		SPADE \cite{park2019semantic} & 0.0389 & 26.50 & 0.8903 & 0.0854 & 0.4812 & 27.79 & 0.8928 & 0.2864 \\
    	\hline
    		Pix2pixHD \cite{wang2018high} & 0.0430 & 26.63 & 0.8893 & 0.0840 & 0.4283 & 27.82 & 0.8910 & 0.2856 \\
    	\hline
        	Pix2pix \cite{isola2017image} & 0.0611 & 26.56 & 0.8870 & 0.0913 & 8.4077 & 26.56 & 0.8855 & 0.3301 \\
    	\hline
	\end{tabular}
\label{table2}
\end{table*}

\noindent \textbf{Evaluation metrics.} To accurately assess model performance, we utilize both image quality metrics and medical imaging semantic segmentation metrics:

Four image quality metrics are considered in this study: Fréchet inception distance (FID) \cite{heusel2017gans}, peak-signal-to-noise ratio (PSNR) \cite{hore2010image}, structural similarity index measure (SSIM) \cite{hore2010image}, and root mean square error (RMSE) \cite{zhu2019sean}. FID measures the similarity of the distributions of real and synthesized images using a deep-learning model. PSNR and SSIM are the most widely used metrics when evaluating the performance of image restoration and reconstruction methods. The former represents the ratio between the maximum possible intensity of a signal and the intensity of corrupting noise, while the latter reflects the structural similarity between two images.

Three semantic segmentation metrics for medical imaging are used in this experiment: the dice score (Dice), sensitivity (Sen), and specificity (Spec) \cite{fenster2006evaluation,milletari2016v}. The dice score evaluates the area of overlap between a prediction and the ground truth, while sensitivity and specificity are two statistical metrics for the performance of binary medical image segmentation tasks. The former measures the percentage of actual positive pixels that are correctly predicted to be positive, while the latter measures the proportion of actual negative pixels that are correctly predicted to be negative. These three metrics are employed for semantic segmentation based on the assumption that, if the quality of the synthesized images is high enough, excellent segmentation performance can be achieved when using the synthesized images as input.

\noindent \textbf{Implementation details.} We transform all of the CT slices into gray-scale images on a Hounsfield unit (HU) scale [- 600,1500]. The sizes of the images and segmentation maps are then rescaled from $630\times630$ to $512\times512$. All of the image synthesis methods are trained with 20 epochs, with a learning rate that is maintained at 0.0002 for the first 10 epochs before linearly decaying to zero over the following ten epochs. Global-local generator $G$ and multi-resolution discriminator $D$ are trained using an Adam optimizer with parameters $\beta_1=0.5$ and $\beta_2=0.999$. The feature matching loss weight $\lambda$ is set at 10. The batch size used to train the proposed method is 16. All of the experiments are run in an Ubuntu 18.04 environment using an Intel i7 9700k CPU and two GeForce RTX Titan graphics cards (48 GB VRAM).

\begin{table*}[]
	\centering
	\caption{EXPERIMENTAL RESULTS FOR CT IMAGES USING SEMANTIC SEGMENTATION METHODS \hspace{\textwidth} (REPLACING REAL DATA WITH DIFFERENT PROPORTIONS OF SYNTHETIC DATA) \hspace{\textwidth} (The best evaluation score is marked in bold. $\uparrow$ means higher number is better, and $\downarrow$ indicates lower number is better. Ratio means replacing certain proportion of synthetic data. $\varepsilon$ represents a small positive quantity which is smaller than $1e^{-5}$. 50\%$^{(1)}$, 50\%$^{(2)}$, 50\%$^{(3)}$, 50\%$^{(4)}$ represent the synthetic data are from SEAN \cite{zhu2019sean}, SPADE \cite{park2019semantic}, Pix2pixHD \cite{wang2018high} and Pix2pix \cite{isola2017image}, respectively.}
	\begin{tabular}{|c|c|c|c|c|c|c|c|c|c|}
		\hline
            Focus & \multicolumn{3}{|c|}{Ground-glass opacity} & \multicolumn{3}{|c|}{Consolidation} &
            \multicolumn{3}{|c|}{Infection}\\ 
		\hline
		    Ratio & Dice (\%, $\uparrow$) & Sen (\%, $\uparrow$) & Spec (\%, $\uparrow$) & Dice (\%, $\uparrow$) & Sen (\%, $\uparrow$) & Spec (\%, $\uparrow$) & Dice (\%, $\uparrow$) & Sen (\%, $\uparrow$) & Spec (\%, $\uparrow$) \\
		\hline
		    0\% & \textbf{87.55$\pm$0.20} & \textbf{86.84$\pm$0.31} & 99.82$\pm$0.01 & 84.88$\pm$0.33 & 82.80$\pm$0.51 & 99.96$\pm\varepsilon$ & \textbf{89.57$\pm$0.18} & \textbf{88.58$\pm$0.23} & 99.82$\pm$0.01 \\
    	\hline
    	    10\% & 87.34$\pm$0.27 & 85.08$\pm$0.45 & \textbf{99.85$\pm$0.01} & 85.91$\pm$0.44 & 84.23$\pm$0.55 & 99.96$\pm\varepsilon$ & 89.35$\pm$0.32 & 87.14$\pm$0.39 & \textbf{99.85$\pm$0.01} \\
    	\hline
        	20\% & 84.22$\pm$0.36 & 83.38$\pm$0.41 & 99.77$\pm$0.01 & 84.30$\pm$0.19 & 83.60$\pm$0.33 & 99.95$\pm\varepsilon$ & 86.67$\pm$0.36 & 85.83$\pm$0.27 & 99.76$\pm$0.01 \\
    	\hline
    	    30\% & 87.43$\pm$0.35 & 85.73$\pm$0.41 & 99.84$\pm$0.01 & \textbf{86.01$\pm$0.21} & \textbf{87.14$\pm$0.21} & 99.95$\pm\varepsilon$ & 89.32$\pm$0.22 & 88.11$\pm$0.30 & 99.82$\pm$0.01 \\
    	\hline
            40\% & 87.07$\pm$0.27 & 86.70$\pm$0.29 & 99.80$\pm$0.01 & 85.81$\pm$0.24 & 81.94$\pm$0.36 & \textbf{99.97$\pm\varepsilon$} & 88.92$\pm$0.20 & 87.90$\pm$0.30 & 99.81$\pm$0.01 \\
    	\hline
    	    50\% & 86.98$\pm$0.35 & 86.46$\pm$0.40 & 99.80$\pm$0.01 & 85.58$\pm$0.23 & 82.18$\pm$0.36 & \textbf{99.97$\pm\varepsilon$} & 89.19$\pm$0.24 & 88.12$\pm$0.37 & 99.82$\pm$0.01 \\
    	\hline
    	    50\%$^{(1)}$ & 85.23$\pm$0.40 & 84.99$\pm$0.31 & 99.80$\pm$0.01 & 83.54$\pm$0.22 & 83.66$\pm$0.19 & \textbf{99.97$\pm\varepsilon$} & 86.04$\pm$0.29 & 86.00$\pm$0.39 & 99.82$\pm$0.01 \\
    	\hline
    	    50\%$^{(2)}$ & 83.04$\pm$0.28 & 81.56$\pm$0.24 & 99.80$\pm$0.01 & 81.89$\pm$0.20 & 81.50$\pm$0.30 & 99.96$\pm\varepsilon$ & 85.99$\pm$0.21 & 84.05$\pm$0.15 & 99.81$\pm$0.01 \\
    	\hline
    	    50\%$^{(3)}$ & 81.24$\pm$0.37 & 79.53$\pm$0.29 & 99.79$\pm$0.01 & 80.20$\pm$0.44 & 78.14$\pm$0.46 & 99.96$\pm\varepsilon$ & 83.22$\pm$0.45 & 83.01$\pm$0.48 & 99.80$\pm$0.01 \\
    	\hline
    	    50\%$^{(4)}$ & 75.33$\pm$0.24 & 71.02$\pm$0.38 & 99.75$\pm$0.01 & 72.01$\pm$0.25 & 70.55$\pm$0.21 & 99.95$\pm\varepsilon$ & 79.10$\pm$0.25 & 78.89$\pm$0.39 & 99.77$\pm$0.01 \\
    	\hline
	\end{tabular}
\label{table3}
\end{table*}

\begin{table*}[]
	\centering
	\caption{EXPERIMENTAL RESULTS FOR CT IMAGES USING SEMANTIC SEGMENTATION METHODS \hspace{\textwidth} (ADDING SYNTHETIC DATA WITH DIFFERENT PROPORTIONS) \hspace{\textwidth} (The best evaluation score is marked in bold. $\uparrow$ means higher number is better, and $\downarrow$ indicates lower number is better. Ratio means adding certain proportion of synthetic data. $\varepsilon$ represents a small positive quantity which is smaller than $1e^{-5}$.)}
	\begin{tabular}{|c|c|c|c|c|c|c|c|c|c|}
		\hline
            Focus & \multicolumn{3}{|c|}{Ground-glass opacity} & \multicolumn{3}{|c|}{Consolidation} &
            \multicolumn{3}{|c|}{Infection}\\ 
		\hline
		    Ratio & Dice (\%, $\uparrow$) & Sen (\%, $\uparrow$) & Spec (\%, $\uparrow$) & Dice (\%, $\uparrow$) & Sen (\%, $\uparrow$) & Spec (\%, $\uparrow$) & Dice (\%, $\uparrow$) & Sen (\%, $\uparrow$) & Spec (\%, $\uparrow$) \\
		\hline
		    0\% & 87.55$\pm$0.20 & 86.84$\pm$0.31 & 99.82$\pm$0.01 & 84.88$\pm$0.33 & 82.80$\pm$0.51 & \textbf{99.96$\pm\varepsilon$} & 89.57$\pm$0.18 & 88.58$\pm$0.23 & 99.82$\pm$0.01 \\
    	\hline
    	    10\% & 87.65$\pm$0.40 & 85.82$\pm$0.32 & \textbf{99.84$\pm$0.01} & 86.12$\pm$0.33 & 86.02$\pm$0.63 & 99.95$\pm\varepsilon$ & 89.67$\pm$0.31 & 88.11$\pm$0.32 & \textbf{99.84$\pm$0.01} \\
    	\hline
        	20\% & 87.87$\pm$0.34 & 87.67$\pm$0.28 & 99.81$\pm$0.01 & 85.52$\pm$0.34 & 84.16$\pm$0.53 & \textbf{99.96$\pm\varepsilon$} & 89.87$\pm$0.12 & 89.44$\pm$0.20 & 99.81$\pm$0.01 \\
    	\hline
    	    30\% & 87.99$\pm$0.36 & 87.25$\pm$0.36 & 99.82$\pm$0.01 & 86.33$\pm$0.30 & 86.38$\pm$0.51 & 99.95$\pm\varepsilon$ & 89.78$\pm$0.22 & 89.17$\pm$0.27 & 99.82$\pm$0.01 \\
    	\hline
            40\% & \textbf{88.33$\pm$0.22} & \textbf{88.71$\pm$0.32} & 99.81$\pm$0.01 & \textbf{87.25$\pm$0.28} & 86.30$\pm$0.38 & \textbf{99.96$\pm\varepsilon$} & \textbf{90.19$\pm$0.17} & \textbf{90.34$\pm$0.31} & 99.81$\pm\varepsilon$ \\
    	\hline
    	    50\% & 88.16$\pm$0.30 & 86.88$\pm$0.01 & \textbf{99.84$\pm$0.01} & 87.09$\pm$0.34 & \textbf{86.54$\pm$0.47} & \textbf{99.96$\pm\varepsilon$} & 90.06$\pm$0.30 & 88.88$\pm$0.18 & 99.83$\pm$0.01 \\
    	\hline
	\end{tabular}
\label{table4}
\end{table*}

\subsection{Quantitative results}
The performance of the proposed method is assessed according to both image quality and medical imaging semantic segmentation.

\subsubsection{Image quality evaluation} 
In this study, common image quality metrics are employed to assess the synthesis performance of the proposed method and four other state-of-the-art image synthesis methods: SEAN \cite{zhu2019sean}, SPADE \cite{park2019semantic}, Pix2pixHD \cite{wang2018high}, and Pix2pix \cite{isola2017image}. We evaluate image quality for two synthetic image categories: complete and lung-only images. The complete images are those CT images generated by merging a synthesized lung CT image with its corresponding non-lung CT image. The evaluation results are presented in Table \ref{table2}.

The proposed method outperforms other state-of-the-art methods based on the four image quality metrics for both the complete and lung-only images. Due to the design of the global-local generator and multi-resolution discriminator, the proposed model can generate realistic lung CT images for COVID-19 with a complete global structure and fine local details and maintain a relatively high signal-to-noise ratio. Thus, the proposed method can achieve state-of-the-art image synthesis results based on image quality.

\subsubsection{Medical imaging semantic segmentation evaluation}
To evaluate the reconstruction capability of the proposed method, we utilize Unet, a common medical imaging semantic segmentation approach \cite{ronneberger2015u}. We first train the Unet model on a mix of synthetic and real CT images. The training set of this task consists of real and synthetic data from the test set of the image synthesis tasks, and the test set here we use the training set of the image synthesis task. 

This evaluation consists of two independent experiments: (1) keeping the total number of images the same while replacing the real data with synthesized data from a proportion of 0\% to 50\% in steps of 10\% and (2) keeping the number of real images the same and adding a certain proportion of synthetic images from 0\% to 50\% in steps of 10\%. The first experiment evaluates how similar the synthetic and real data are and the second evaluates the image synthesis potential of the synthetic data. We consider three categories in the assessment: ground-glass opacity, consolidation, and infection (which considers both ground-glass opacity and consolidation). The evaluation results for the two experiments are presented in Table \ref{table3} and Table \ref{table4}, respectively. The pre-trained Unet model is then tested with a fixed real CT image dataset. 10,220 images from the test set are divide equally into 10 folds, the evaluation results are reported with the format as \textit{MEAN} $\pm$ \textit{95\% CONFIDENCE INTERVAL} among above folds.

In Table III, we describe the experimental results of different replacing ratios of synthetic data. We can obtain the best performance when using pure real data as a training set. By replacing the real data with a ratio of synthetic data, the semantic segmentation performance of Unet does not decrease and stay at a stable level. By replacing real data with 30\% synthetic data, the Unet obtains the best performance on the Spec metric for ground-glass opacity focus, also it gets the best performance on Dice and Sen metrics for consolidation focus. The experimental results from Table III show that synthetic CT images are similar to real CT images. They are realistic enough even replacing the real data with a large ratio of synthetic data, the semantic segmentation performance of Unet still seems promising. Besides, we also demonstrate the performance comparison with other state-of-the-art image synthesizers in Table \ref{table3}. Under the condition of replacing real data with 50\% synthetic data which is generated by four different competitors, the proposed method shows the competitive performance on the semantic segmentation tasks. 

Table \ref{table3} presents the experimental results for different replacement ratios for the synthetic data. We obtain the best performance when using pure real data as the training set. By replacing the real data with a proportion of synthetic data, the semantic segmentation performance of Unet does not decrease, but rather remains stable. By replacing real data with 30\% synthetic data, Unet obtains the best performance for the Spec metric for ground-glass opacity and for the Dice and Sen metrics for consolidation. The experimental results thus indicate that the synthetic CT images are similar to real CT images. They are sufficiently realistic for semantic segmentation with Unet to be successful even when real data is replaced with a large proportion of synthetic data.

Table \ref{table4} presents the semantic segmentation results when a certain proportion of extra synthetic data is added to the real data. The best performance is obtained when adding 40\% synthetic data. Overall, the results indicate that the synthetic CT images are sufficiently diverse and realistic, meaning that they have the potential to be utilized in image synthesis to improve the dataset quality for deep-learning-based COVID-19 diagnosis.

\subsection{Qualitative results}

\begin{figure*}[!ht]
\centering
\includegraphics[width=18cm]{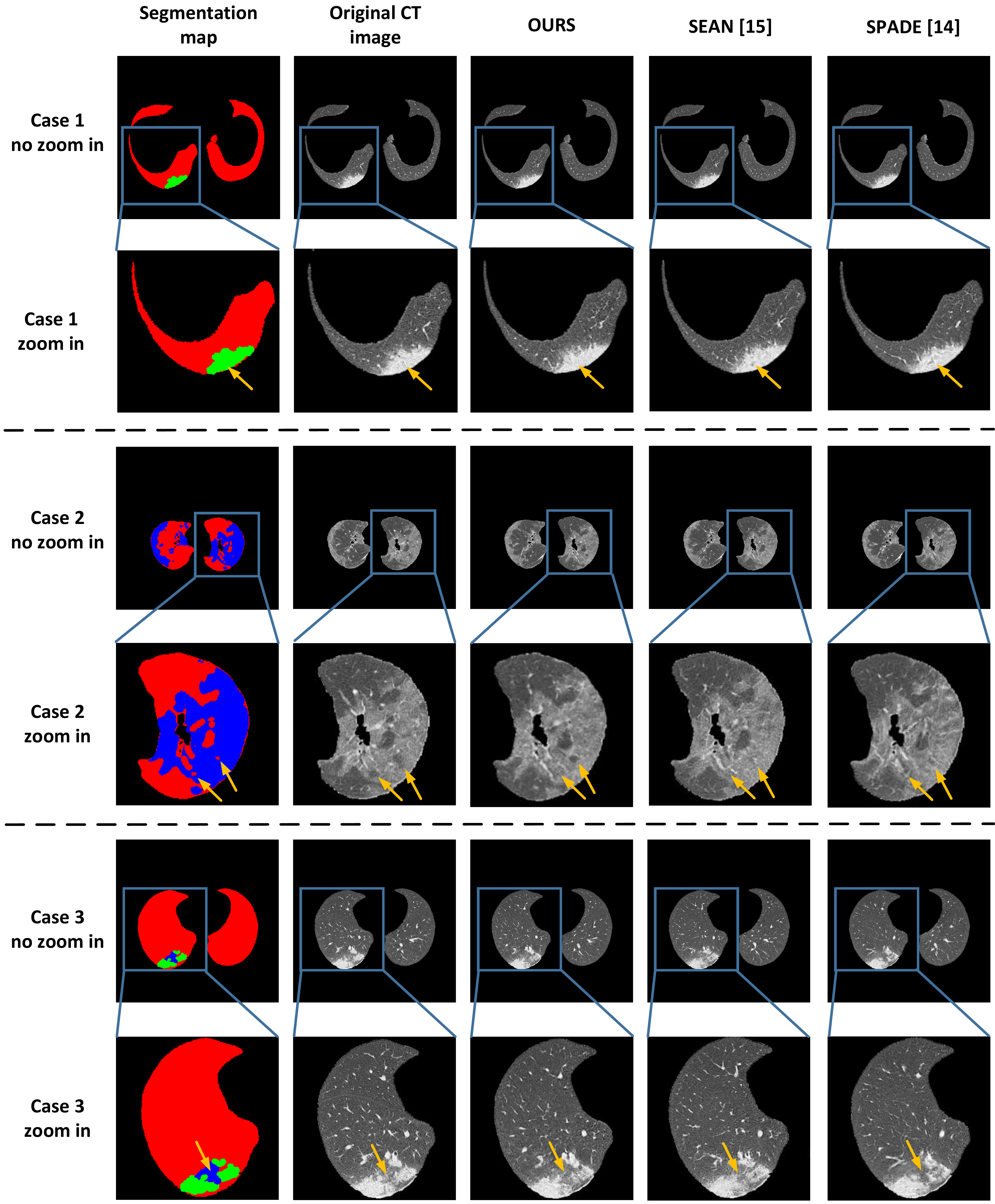}
\caption{Synthetic lung CT images generated by the proposed method and the other two competitive state-of-the-art image synthesis approaches. The first column shows the segmentation map including the lung (red), ground-glass opacity (blue), and consolidation (green) areas. The second column shows the original CT image. The third, fourth, fifth columns show the synthetic samples which are generated by the proposed method, SEAN \cite{zhu2019sean} and SPADE \cite{park2019semantic} in order. Each case is presented with zoom in order to show more details, and the yellow arrows point out the special area which is described in the main text.}
\label{fig6} % this gives the figure a unique name that you can refer to in the main text \ref{fig:...}
\end{figure*}

\begin{figure*}[!ht]
\centering
\includegraphics[width=15cm]{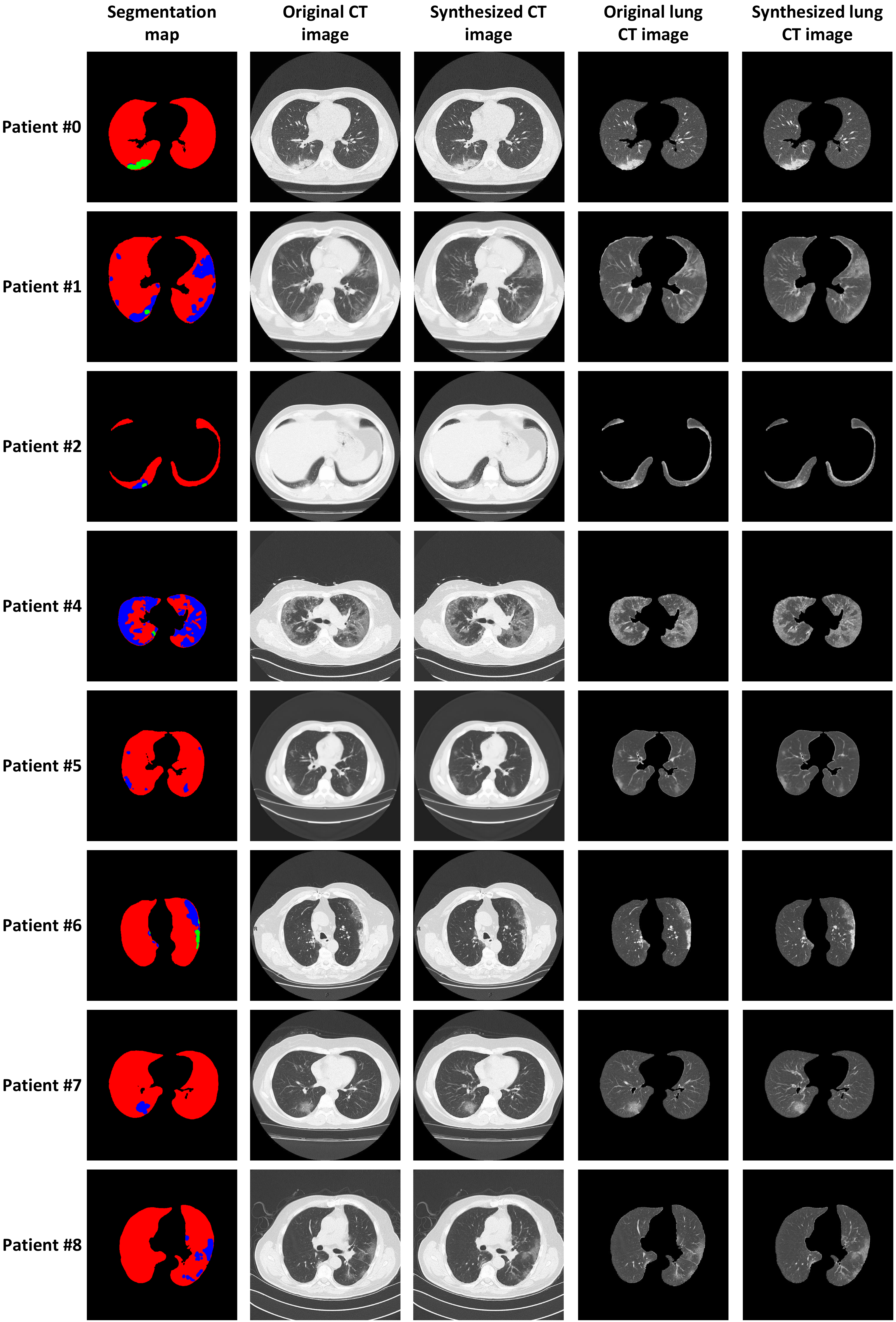}
\caption{Synthetic lung CT images generated by the proposed method. Eight samples are selected, each from an individual patient. The first column shows the segmentation map including the lung (red), ground-glass opacity (blue), and consolidation (green) areas. The second and third columns show the original and synthetic CT images, respectively. The synthetic CT images here merge the synthetic lung CT image and the corresponding real non-lung area. The fourth and fifth columns depict CT images for the original lung and synthesized CT images, respectively.}
\label{fig7} % this gives the figure a unique name that you can refer to in the main text \ref{fig:...}
\end{figure*}

To intuitively demonstrate synthetic results and easily compare them with the results from other state-of-the-art image synthesis methods, we show the synthetic examples in both Figure \ref{fig6} and Figure \ref{fig7} in this subsection.

The synthetic images from three individual cases are compared in Figure \ref{fig6}. The first case shows that a consolidation infection area locates on the lower left of CT image. By comparing the synthetic results from the proposed method, SEAN \cite{zhu2019sean} and SPADE \cite{park2019semantic}, the infection area remains the original structure and texture in the result which is generated by the proposed method, however, we found that in the results of SEAN and SPADE, there some unnatural artifacts (holes) are generated in the position that yellow arrow points out. For the second case, a large area of ground-glass infection is detected, the results of SPADE and SEAN ignore some small lung area in the middle of the infection area, but the proposed method can still reflect above small lung area correctly. In the final case, it contains both two categories of infection area: consolidation and ground-glass opacity, and the ground-glass opacity is surrounded by the consolidation area. If we focus on the surrounded area, we can found out that the boundary of two infection area is not clear in the synthetic image of SEAN, and the ground-glass area are mistakenly generated as lung area in the synthesized image of SPADE. The result of the proposed method in case 3 shows that it has the ability to handle this complex situation and produce realistic synthetic CT images with high image quality. 

We present some synthetic examples that are generated by the proposed method in Figure \ref{fig7}. We select one example for each patient (8 samples from 9 patients; patient \#3 is skipped because the segmentation maps were miss-labeled). For Patient \#0, the consolidation area is located at the bottom of the lung area; the synthetic image shows a sharp and high-contrast consolidation area that can be easily distinguished from the surrounding non-lung region. The slides for Patients \#1 and \#4 have a similarity in that the lung area contains widespread ground-glass opacity. Consolidation is sporadically located within this ground-glass opacity. The small consolidation area can be easily identified due to the clear boundary between the two infection areas. Patient \#6 shows ground-glass opacity and consolidation that are distant from each other. The results thus illustrate that the proposed method can handle the two types of infection areas together in a single lung CT image. The CT slides of Patients \#5, \#7, and \#8 show the simplest cases, with only a single category of infection (ground- glass opacity). The experimental results thus indicate that realistic ground-glass opacity can be obtained using the proposed method.

\subsection{Discussion}

\begin{table}[htbp] 
 \centering
 \caption{Ablation study of various proposed model structure} 
 \begin{tabular}{cccc} 
  \toprule 
  Method & FID ($\downarrow$) & PSNR ($\uparrow$) & Dice (\%, $\uparrow$) \\ 
  \midrule 
  Ours & 0.0327 & 26.89 & 89.19$\pm$0.24 \\
  \midrule 
 w/o DESUM & 0.0395 & 26.77 & 84.60$\pm$0.39 \\
 using $F_{global}$ & 0.0355 & 26.70 & 87.88$\pm$0.29 \\
 Fixed $\alpha=0.5$ & 0.0380 & 26.51 & 85.87$\pm$0.31 \\
 \midrule 
 w/o DFM & 0.0381 & 26.62 & 86.10$\pm$0.33 \\
 using $D_{2}^{i}$ & 0.0340 & 26.75 & 88.84$\pm$0.21 \\
 Fixed $\beta=0.5$ & 0.0404 & 26.71 & 83.50$\pm$0.40 \\
 \midrule 
 D=1 & 0.0579 & 26.55 & 79.44$\pm$0.80 \\
 D=3 & 0.0330 & 26.80 & 88.93$\pm$0.26 \\
 \midrule 
 G=1 & 0.0604 & 26.51 & 76.03$\pm$0.52 \\
 G=3 & 0.0325 & 26.85 & 89.01$\pm$0.20 \\
  \bottomrule 
 \end{tabular} 
\label{table5}
\end{table}

In order to further discuss the efficiency of DESUM process and DFM process, and justify an optimal structure of cGAN, we follow the experimental settings of the image quality evaluation and medical imaging semantic segmentation evaluation in sub-section B. Experimental results are shown in Table \ref{table5}.

\subsubsection{Dynamic element-wise sum process (DESUM)} 

In the second part of Table \ref{table5}, we evaluate the performance of three variations that are related to DESUM. Without DESUM, using $F_{global}$, one fixed $\alpha$ number. The performance shows that DESUM can effectively improve the image quality and segmentation performance. The intermediate feature $F_{global}$ offers less information than $F_{global}$ does, which influences the efficiency of DESUM. Using a fixed number of $\alpha$ can not help to boost the performance, sometimes it may reduce the performance since a fixed weight is not suitable for diverse COVID-19 CT data.

\subsubsection{Dynamic feature matching process (DFM)} 

In this sub-section, we discuss the evaluation of DFM. As shown in the third part of Table \ref{table5}, we compared the performance of the various (a) without DFM, (b) using the intermediate feature from $D_1$, (c) using a fixed $\beta$ number. The evaluation results show that DFM can help to train discriminator stably and improve the performance with multiple metrics. Moreover, the intermediate feature from $D_1$ contains many more details that can help the DFM process to weight correctly. The results also tell that a dynamic weight of $\beta$ is critical for training our multi-resolution discriminator.

\subsubsection{Fine-tuning level optimization} 

We also investigated some potential fine-tuning optimizations which are the number of generators and the number of discriminators. In the last two parts of Table \ref{table5}, we found that the number of generators and discriminators is the important hyper-parameters of COVID-19 CT image synthesis task. A proper number of generator and discriminator can not only avoid the model from overfitting with details from multiple resolution sources but also improve the training efficiency and stability. Experimental results show that the performance can benefit from the dual structure of both generator and discriminator the most, because this dual structure can trade-off well between performance and efficiency.

\section{Conclusion and future study}
In this paper, we proposed a cGAN-based COVID-19 CT image synthesis method that can generate realistic CT images that included two main infection types; ground-glass opacity and consolidation. The proposed method takes the semantic segmentation map of a corresponding lung CT image, and the cGAN structure learns the characteristics and information of the CT image. A global-local generator and a multi-resolution discriminator are employed to effectively balance global information with local details in the CT image. The experimental results have shown that the proposed method was able to generate realistic synthetic CT images and achieve state-of-the-art performance in terms of image quality when compared with common image synthesis approaches. In addition, the evaluation results for semantic segmentation performance demonstrated that the high image quality and fidelity of the synthetic CT images enable their use in image synthesis for COVID-19 diagnosis using AI models. For future research, the authors plan to fully utilize high-quality synthetic COVID-19 CT images to improve specific computer vision approaches that can help in the fight against COVID-19, such as lung CT image semantic segmentation and rapid lung CT image-based COVID-19 diagnosis.

\ifCLASSOPTIONcaptionsoff
  \newpage
\fi

\bibliographystyle{IEEEtran}
\bibliography{IEEEtran}{}

\end{document}